\documentclass[sigconf]{acmart}

\usepackage{booktabs} % For formal tables

\usepackage{eurosym}

\usepackage{xcolor}
\definecolor{myblue}{RGB}{31, 73, 125} 

\usepackage{changepage}

\setcopyright{rightsretained}
\acmDOI{}

\acmISBN{}

\acmConference{}{}{}
\acmYear{2018}
\copyrightyear{2018}

\acmArticle{4}
\acmPrice{15.00}

\begin{document}
\title[Quantum Internet: from Communication to Distributed Computing!]{Quantum Internet:\\from Communication to Distributed Computing!}
%\titlenote{Produces the permission block, and copyright information}
%\subtitle{Invited Paper}
%\subtitlenote{The full version of the author's guide is available as \texttt{acmart.pdf} document}

\author{Marcello Caleffi}
\affiliation{
	\institution{University of Naples \textit{Federico II}}
	\streetaddress{Via Claudio 21}
	\city{Naples}
	\state{ITALY}
	\postcode{80125}
}
%\additionalaffiliation{
%	\institution{National Inter-University Consortium for Telecommunications (CNIT)}
%	\streetaddress{Via Cintia 4 }
%	\city{Naples}
%	\state{ITALY}
%	\postcode{80126}
%}
\email{marcello.caleffi@unina.it}

\author{Angela Sara Cacciapuoti}
%\authornote{Dr.~Trovato insisted his name be first.}
\affiliation{
	\institution{University of Naples \textit{Federico II}}
	\streetaddress{Via Claudio 21}
	\city{Naples}
	\state{ITALY}
	\postcode{80125}
}
%\additionalaffiliation{
%	\institution{National Inter-University Consortium for Telecommunications (CNIT)}
%	\streetaddress{Via Cintia 4 }
%	\city{Naples}
%	\state{ITALY}
%	\postcode{80126}
%}
\email{angelasara.cacciapuoti@unina.it}

\author{Giuseppe Bianchi}
%\authornote{Dr.~Trovato insisted his name be first.}
\affiliation{
	\institution{University of Roma \textit{Tor Vergata}}
	\streetaddress{Via del Politecnico 1}
	\city{Roma}
	\state{ITALY}
	\postcode{00133}
}
%\additionalaffiliation{
%	\institution{National Inter-University Consortium for Telecommunications (CNIT)}
%	\streetaddress{Via Cintia 4 }
%	\city{Naples}
%	\state{ITALY}
%	\postcode{80126}
%}
\email{giuseppe.bianchi@uniroma2.it}

%\ref{?}newcommand{\shortauthors}{B. Trovato et al.}

% -------------------------------------------------------------------------------------------------------------------------------------------------------------------------------
% abstract
% -------------------------------------------------------------------------------------------------------------------------------------------------------------------------------
\begin{abstract}
In this invited paper, the authors discuss the exponential computing speed-up achievable by interconnecting quantum computers through a quantum internet. They also identify key future research challenges and open problems for quantum internet design and deployment.
\end{abstract}

% -------------------------------------------------------------------------------------------------------------------------------------------------------------------------------
% css concepts
% -------------------------------------------------------------------------------------------------------------------------------------------------------------------------------


% -------------------------------------------------------------------------------------------------------------------------------------------------------------------------------
% key words
% -------------------------------------------------------------------------------------------------------------------------------------------------------------------------------

\maketitle

% -------------------------------------------------------------------------------------------------------------------------------------------------------------------------------
% Section 1
% -------------------------------------------------------------------------------------------------------------------------------------------------------------------------------
\section{Introduction}

Quantum computing is not a novel concept: it was proposed in the early eighties \cite{Fey-82} as a disruptive paradigm to solve challenging problems exponentially faster than classical computing can. In 1994, Shor proved its disruptive potential for integer factorization \cite{Sho-94}, which constitutes one of the most widely adopted algorithms for securing communications over the Internet. Since then, quantum computing attracted a great academic interest, mainly from the physics communities.

\begin{figure}[t]
	\centering
	\includegraphics[width=1\columnwidth]{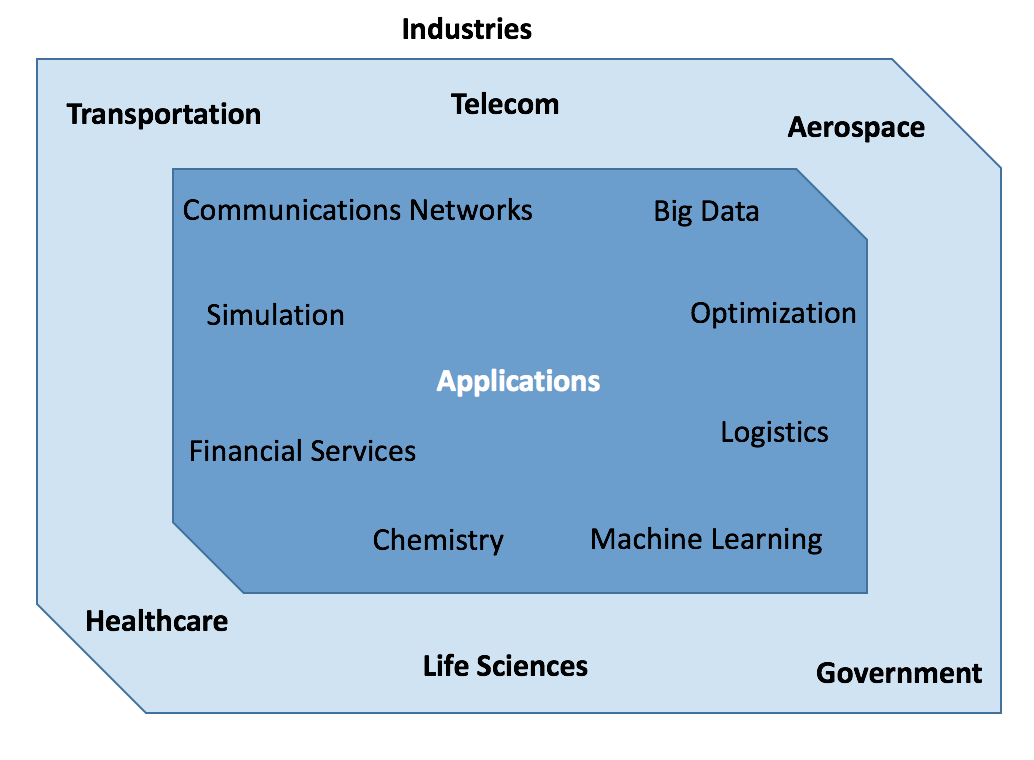}
	\caption{Quantum Computing Industries and Applications.}
	\label{fig:01}
\end{figure}

Nowadays, after over twenty years of \textit{pure science phase}, the research on quantum computing is finally reaching the \textit{engineering phase}, getting out of the labs and into business reality. 

In fact, the development of quantum computers is experiencing a major boost, since tech giants entered the \textit{quantum race}. In November 2017 IBM built and tested a $50$-qubits processor \cite{IBM-17}, in March 2018 Google announced a $72$-qubits processor \cite{Kel-18}, and other big players, like Intel and Alibaba, are actively working on double-digit-qubits proof-of-concepts. Meanwhile, in April 2017 the European Commission launched a ten-years 1~\euro-billion flagship project to boost European quantum technologies research \cite{Gib-17}, and in June 2017 China successfully tested a 1200km quantum communication between satellite Micius and ground stations \cite{LiaCai-17}.

Such a widespread excitement about quantum computing is not surprising, given its potential to completely change markets and industries -- such as commerce, intelligence, military affairs -- as shown by the plethora of applications in Figure~\ref{fig:01}. Potentially, a quantum computer can tackle classes of problems that choke conventional machines, such as molecular and chemical reaction simulations \cite{Bou-17}, optimization in manufacturing and supply chains, financial modeling, machine learning and enhanced security.

For instance, with reference to our previous integer factorization example, Shor's quantum algorithm exhibits polynomial complexity. This means that a $2048$bit random integer can be factorized in few minutes (or hours) by using a quantum computer, but it takes billions of years -- more than the age of universe -- by using a classical computer \cite{VanKohLad-12}.

But the complete overtaking of classical computing is expected to require hundreds, thousands of fault-tolerant qubits embedded in a single quantum chip \cite{Bou-17}. However, quantum technologies are still far away from this ambitious goal, and it is impossible to predict if this revolution will take years or decades.

In fact, so far, although the quantum chips storing the qubits are quite small, with dimensions comparable to classical chips, they require to be confined into specialized laboratories hosting the bulked equipment -- such as large sub-absolute-zero cooling systems -- necessary to preserve the coherence of the quantum states. And the challenge for controlling, interconnecting, and preserving the qubits gets harder and harder as the number of qubits increases.

All that given, it might sound like the dawn of quantum computing revolution hasn't arrived. Yet, the market for quantum computing is forecast to worth more than $10$~\$-billion by 2024 \cite{HSRC-18}, and several technology firms already begun to prepare their business for the era of quantum computing \cite{Acc-17}.

% -------------------------------------------------------------------------------------------------------------------------------------------------------------------------------
% Section 2
% -------------------------------------------------------------------------------------------------------------------------------------------------------------------------------
\section{Quantum Computing Background}
\label{sec:2}

\begin{figure}[t]
	\centering
	\includegraphics[width=1\columnwidth]{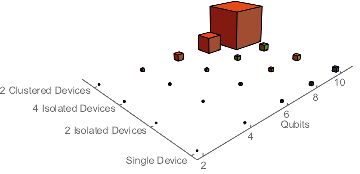}
	\caption{Quantum cloud computing speed-up: even in case of small quantum devices, two clustered devices can provide an exponential speedup with respect to isolated devices. The volume of the cube represents the computation power. }
	\label{fig:speed}
\end{figure}

The power of quantum computing lies in its foundation on \textcolor{myblue}{\textbf{quantum mechanics}}, the deepest explanation of the reality surrounding us.

Qubits are the building blocks of quantum computation. Qubits can be realized with different technologies -- electron spin, light polarization, super-conducting circuits \cite{VanDev-16} -- but the principles of quantum mechanics hold independently of the underlying technology.

While a classical bit can be in one state -- either 0 or 1 -- a qubit can be in a \textcolor{myblue}{\textbf{superposition}} of two states 0 and 1 simultaneously. 2 qubits can exist in a superposition of 4 states. With 3 qubits, the states become 8 and, with 4 qubits, the states become 16. Thanks to the superposition principle, \textcolor{myblue}{\textbf{the power of quantum computing grows exponentially as more qubits are added}}.

Quantum states are fragile. Any interaction with the environment irreversibly affects any quantum state, causing a loss of its quantum properties in a process called \textcolor{myblue}{\textbf{decoherence}}. The classical strategy -- storing redundant copies of the fragile data -- is not a solution. The \textcolor{myblue}{\textbf{no-cloning theorem}}, in fact, prohibits to make a copy of an arbitrary quantum state, and it turns out to be a valuable property for securing communications.

By \textcolor{myblue}{\textbf{measuring a qubit}}, the quantum state is irremediably altered: any superposition probabilistically collapses into a single state. The measuring principle deeply affects computing: clever ways to manipulate quantum states without measuring are essential for quantum algorithm design.

% -------------------------------------------------------------------------------------------------------------------------------------------------------------------------------
% Section 3
% -------------------------------------------------------------------------------------------------------------------------------------------------------------------------------
\section{Quantum Internet}
\label{sec:3}
From the above, it is now clear that the computational power of a quantum device is dictated by the number of qubits that can be embedded and interconnected within. But the state-of-the-art of quantum technologies limits this number to double digits. 

Hence, a question spontaneously arises: \textcolor{myblue}{\textbf{how can we significantly scale up the number of qubits to achieve the quantum supremacy?}}

\vspace{3pt}
\begin{adjustwidth}{18pt}{18pt}
Interconnecting multiple quantum devices via a \textcolor{myblue}{\textbf{quantum internet}}, i.e., through a network able to share quantum states among remote nodes (opposed to the reductive usage of quantum communications ``just'' for securing an otherwise ordinary data transfer), is the answer.
\end{adjustwidth}
\vspace{3pt}

Two isolated $10$-qubit devices can represent $2^{10}$ states each thanks to the superposition principle. Hence, the two isolated devices represent $2 \cdot 2^{10}$ states at once. But if we interconnect these two devices with a quantum internet, the resulting cluster can represent up to $2^{18}$ states, as detailed at the end of the next section, with an exponential computational speed-up as shown in Figure~\ref{fig:speed}.

In this \textcolor{myblue}{\textbf{distributed quantum computing scenario}}, existing data centers are the natural candidates for hosting the specialized quantum computing equipment. And companies and users can access to the quantum computing power as a service via cloud. Indeed, the \textcolor{myblue}{\textbf{quantum cloud}} market is estimated nearly half of the whole 10 billion quantum computing market by 2024 \cite{HSRC-18}. And IBM is already allowing researchers around the world to practice quantum algorithm design through a cloud access to isolated 5-, 16- and 20-qubits quantum devices \cite{Cas-16}.

% -------------------------------------------------------------------------------------------------------------------------------------------------------------------------------
% Section 4
% -------------------------------------------------------------------------------------------------------------------------------------------------------------------------------
\section{Entanglement: the core of Quantum Internet}
\label{sec:4}

\textcolor{myblue}{\textbf{Entanglement}}, defined as a \textit{spooky action at distance} by Einstein, is a property of two quantum particles. The particles exist in a shared state, such that any action on a particle affects instantaneously the other particle as well. This sort of \textit{quantum correlation}, with no counterpart in the classical world, holds even when the particles are far away each other.

Entanglement provides an invaluable tool to transmit qubits without violating the no-copying theorem and the measuring principle. With just local operations and an entangled pair of qubits shared between source and destination, it is possible to ``transmit'' an unknown quantum state between two remotes quantum devices.
	
This process is known as \textcolor{myblue}{\textbf{quantum teleportation}}. Differently from what it is suggested by its name, the quantum teleportation process implies the destruction of the original qubit and the source-side entanglement pair member at the source though a measurement. The original qubit is reconstructed at the destination once the output of the source measurement -- $2$ classical bits -- is received via classical channel.

Stemming from the above description, quantum teleportation is the key strategy to enable distributed quantum computing, which requires to perform operations between qubits physically located on multiple remote quantum devices. Indeed, if we consider for the sake of explanation two remote quantum devices, by devoting at least one qubit at each device for the teleporting process, a virtual quantum device constituted by up to $2n-2$ qubits is obtained. As a consequence, interconnecting multiple quantum devices allows to achieve the  exponential computational speed-up shown in Figure~\ref{fig:speed}.

% -------------------------------------------------------------------------------------------------------------------------------------------------------------------------------
% Section 5
% -------------------------------------------------------------------------------------------------------------------------------------------------------------------------------
\section{Challenges and Open Problems}
\label{sec:5}

\begin{figure}[t]
	\centering
	\includegraphics[width=1\columnwidth]{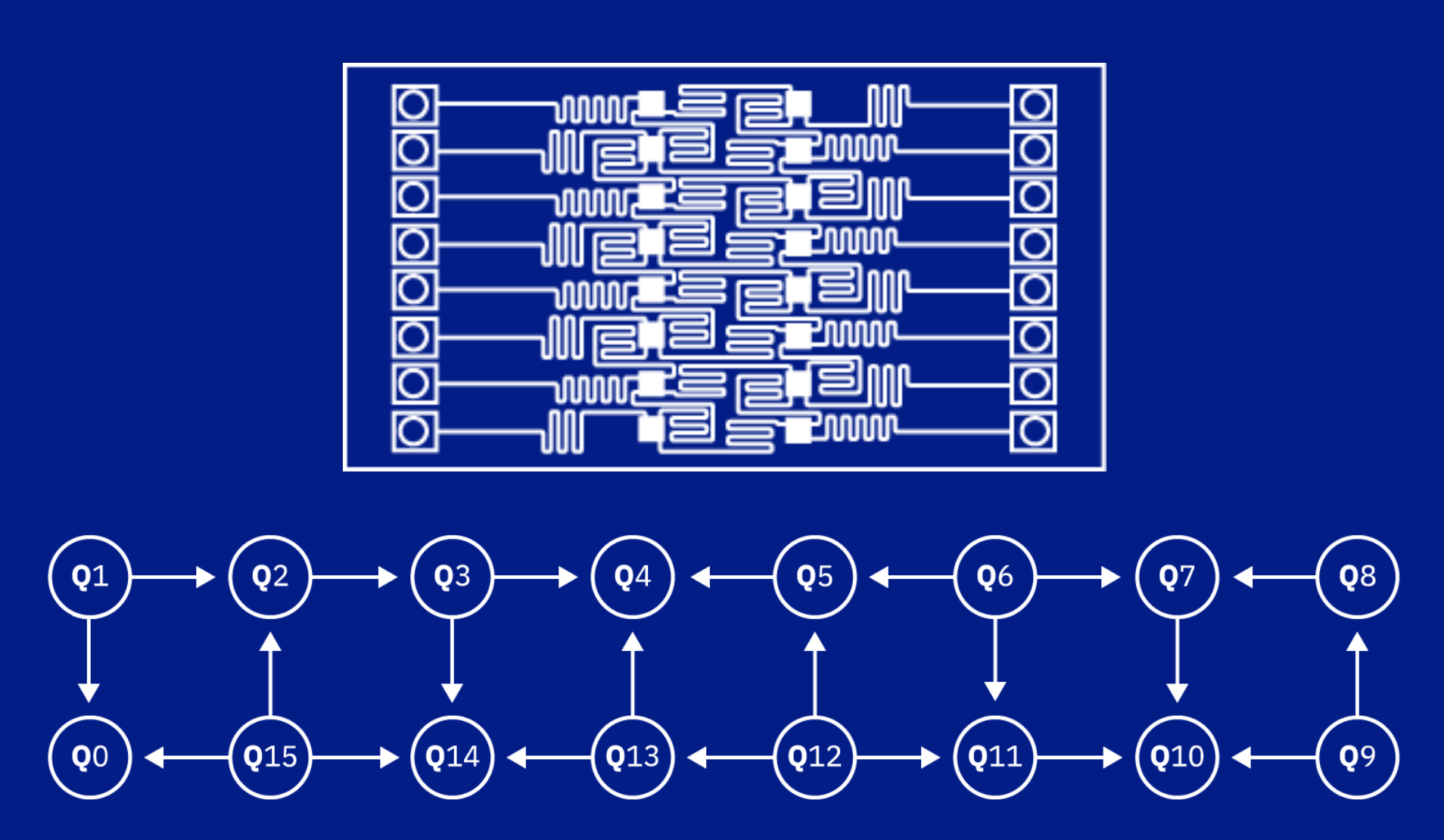}
	\caption{Physical coupling map for the IBM $16$-qubits device. Copyright IBM Q Experience. Qubit $Q_0$ can directly interact with qubits $Q_1$ and $Q_{15}$, but not with qubit $Q_9$, unless the quantum state stored in $Q_0$ is ``transmitted'' to $Q_{10}$ with a sequence of SWAP operations.}
	\label{fig:03}
\end{figure}

The standard computation model of a single quantum device assumes that interactions between arbitrary pairs of physical qubits are available. However, real quantum device architectures have constraints on the physical qubit connectivity \cite{LinMas-17}, as shown in Figure~\ref{fig:03}. These restrictions do not limit in principle the ability to perform arbitrary computations. In fact, swap operations can be used to ``transmit" quantum states between an arbitrary pair of qubits, by using the physical connectivity map. This at the price of longer computation times.

\vspace{3pt}
\begin{adjustwidth}{18pt}{18pt}
\textcolor{myblue}{\textbf{The time overhead induced by swap operations in single-device quantum computing and the time overhead induced by teleporting operations in distributed quantum computing are the two sides of the same coin.}}
\end{adjustwidth}
\vspace{3pt}

They depend on several factors, dictated by the underlying physical hardware/network architecture. They can be controlled by minimizing swapping/teleporting operations with an efficient mapping between the quantum algorithm and the underlying physical architecture. Clearly, this constitutes a key challenge, and indeed IBM launched a quantum computing prize for the swapping operation minimization in January 2018.

Furthermore, the key component underlying the quantum cloud infrastructure is the quantum internet \cite{Cas-18}, which relies on the entanglement through the teleporting process. With the current technology level, photons are considered the natural candidates for generating entanglement between remote qubits. 
And the rationale for this choice lies in the advantages provided by photons for entanglement distribution: weak interaction with the environment, easy control with standard optical components as well as high-speed low-loss transmissions.

The aim of the \textit{flying} qubits (i.e., the photons) is to "transport" the qubit out of the physical quantum device at the sender into the corresponding quantum device at the receiver. 
\vspace{3pt}
\begin{adjustwidth}{18pt}{18pt}
\textcolor{myblue}{\textbf{Hence, a transducer is needed to convert a \textit{matter} qubit, i.e., a qubit for information processing/storage within a computing devices, in a flying qubit, which creates the remote entanglement.}}
\end{adjustwidth}
\vspace{3pt}
However, there exist multiple technologies for realizing a matter qubit (quantum dots, trasmon, ion traps, etc). Each technology is characterized by different pros and cons, as surveyed in \cite{VanDev-16}. As a consequence, the matter-flying interface should be able to face with this technology diversity.

Furthermore, although photons are widely used as information carriers within current internet, the design of a quantum internet requires a major paradigm shift.

\vspace{3pt}
\begin{adjustwidth}{18pt}{18pt}
\textcolor{myblue}{\textbf{The quantum internet is governed by the laws of quantum mechanics. Hence, strange phenomena with no counterpart in the classic reality, such as no-cloning, measurement, entanglement and teleporting, imposes terrific constraints for the network design.}}
\end{adjustwidth}
\vspace{3pt}

Classical network functionalities, ranging from error-control mechanisms -- as ARQ -- to overhead-control strategies -- as caching -- are based on the assumption that you can safety read and copy the information. But this assumption does not hold in a quantum internet, and fundamental advances in the design of each network layer are required \cite{Cal-18}.

\vspace{3pt}
\begin{adjustwidth}{18pt}{18pt}
\textcolor{myblue}{\textbf{In this perspective, quantum networks can not be restricted to a synonymous of specific applications -- e.g. Quantum Key Distribution (QKD) -- as widely done so far. Indeed, for a QKD system, quantum mechanics plays a role only for the creation of the encryption key. The subsequent transmission of the encrypted information is entirely classical. Differently, the Quantum Internet paradigm requires communication networks able to harness entanglement and teleportation for transmitting the fragile quantum information over long distances.}}
\end{adjustwidth}
\vspace{3pt}

We do look forward to contributing to such an exciting research area, which could pave the way for the \textcolor{myblue}{\textbf{Internet of future}} such as Arpanet paved the way for today's internet.

% --------------------------------------------------------------------------------------------------------------------------------------------------------------------------
% Bibliography
% --------------------------------------------------------------------------------------------------------------------------------------------------------------------------
\bibliographystyle{ACM-Reference-Format}
\bibliography{CalCacBia-18}

\end{document}